\documentclass[10pt,journal,compsoc,letterpaper,twoside]{IEEEtran}

\ifCLASSOPTIONcompsoc
  \usepackage[nocompress]{cite}
\else
  \usepackage{cite}
\fi
\ifCLASSINFOpdf
  \usepackage[pdftex]{graphicx}
  \graphicspath{{figures/}{pictures/}{images/}{./}}
  \DeclareGraphicsExtensions{.pdf,.jpeg,.png}
\else
\fi
\usepackage{mathtools,amssymb,amsmath}
\usepackage[basic]{complexity}
\ifCLASSOPTIONcompsoc
 \usepackage[caption=false,font=footnotesize,labelfont=sf,textfont=sf]{subfig}
\else
  \usepackage[caption=false,font=footnotesize]{subfig}
\fi
\usepackage{tabularx,multirow}
\usepackage{color}
\usepackage[usenames,dvipsnames,svgnames,table]{xcolor}
\usepackage{paralist}
\usepackage{hyperref}

\newcommand{\V}{\ensuremath{\mathcal{V}}}
\newcommand{\Sets}{\ensuremath{\mathcal{S}}}
\newcommand{\HG}{\ensuremath{\mathcal{H}}}
\newcommand{\col}{\ensuremath{\mathsf{col}}}
\newcommand{\auc}{\ensuremath{\mathsf{AUC}}}
\newcommand{\comp}[1]{\ensuremath{\Gamma_{#1}}}

\newsavebox{\measurebox}

\newcommand{\revision}[1]{\textcolor{black}{#1}}

\begin{document}
\title{LinSets.zip: Compressing Linear Set Diagrams}

\author{Markus~Wallinger,
        Alexander~Dobler,
        and~Martin~N\"ollenburg%
\IEEEcompsocitemizethanks{\IEEEcompsocthanksitem 
All authors are with the Algorithms and Complexity Group, TU Wien, Vienna, Austria. Email: \{mwallinger, adobler, noellenburg\}@ac.tuwien.ac.at}
}

\markboth{IEEE Transactions on Visualization and Computer Graphics}%
{Wallinger \MakeLowercase{\textit{et al.}}: LinSets.zip: Compressing Linear Set Diagrams}
\IEEEtitleabstractindextext{%
\begin{abstract}

    \revision{
Linear diagrams are used to visualize set systems by depicting set memberships as horizontal line segments in a matrix, where each set is represented as a row and each element as a column. Each such line segment of a set is shown in a contiguous horizontal range of cells of the matrix indicating that the corresponding elements in the columns belong to the set. As each set occupies its own row in the matrix, the total height of the resulting visualization is as large as the number of sets in the instance. Such a linear diagram can be visually sparse and intersecting sets containing the same element might be represented by distant rows. 
To alleviate such undesirable effects, we present LinSets.zip, a new approach that achieves a more space-efficient representation of linear diagrams. 
First, we minimize the total number of gaps in the horizontal segments by reordering columns, a criterion that has been shown to increase readability in linear diagrams.
The main difference of LinSets.zip to linear diagrams is that multiple non-intersecting sets can be positioned in the same row of the matrix. 
Furthermore, we present several different rendering variations for a matrix-based representation that utilize the proposed row compression.
We implemented the different steps of our approach in a visualization pipeline using integer-linear programming, and suitable heuristics aiming at sufficiently fast computations in practice. 
We conducted both a quantitative evaluation 
and 
a small-scale user experiment to compare the effects of compressing linear diagrams.
}

\end{abstract} %

}

\maketitle

\IEEEdisplaynontitleabstractindextext
\IEEEpeerreviewmaketitle

\IEEEraisesectionheading{\section{Introduction}\label{sec:introduction}}
\IEEEPARstart{S}{et} \revision{
systems occur naturally in various use-cases, such as social networks, document analysis, biological data, or more generally, whenever categorical data can be grouped. The visualization of such set systems is crucial in understanding the relationship between elements and sets, sets and sets, or attributes and sets. 
Linear diagrams are a set visualization approach that has been recently proposed~\cite{RodgersSC15}.} In such linear diagrams sets are depicted as one or more line segments in a matrix. Each row represents a single set and each column represents a single element. Each line segment of a set is shown in a contiguous range of cells of the matrix whenever the corresponding elements (columns) belong to the set (row).  
\revision{The focus of linear diagrams is on aggregating membership of individual elements as contiguous segments, thus, focusing on overlapping sets similar to Euler diagrams.}
However, it has been empirically shown in user studies that linear diagrams significantly outperform Euler diagrams on set-theoretic tasks~\cite{ChapmanSRMB14, Gottfried15}. Reasons are that Euler diagrams might not be well-matched or proportional depending on the structure of the represented data and therefore are harder to read or it is even impossible to visualize all occurring relationships between sets~\cite{RodgersZP12}.

\begin{figure*}
\centering
\sbox{\measurebox}{%
  \begin{minipage}[b]{.46\textwidth}
  \subfloat
    [Linear Diagram]
    {\label{fig:figA}\includegraphics[width=\textwidth]{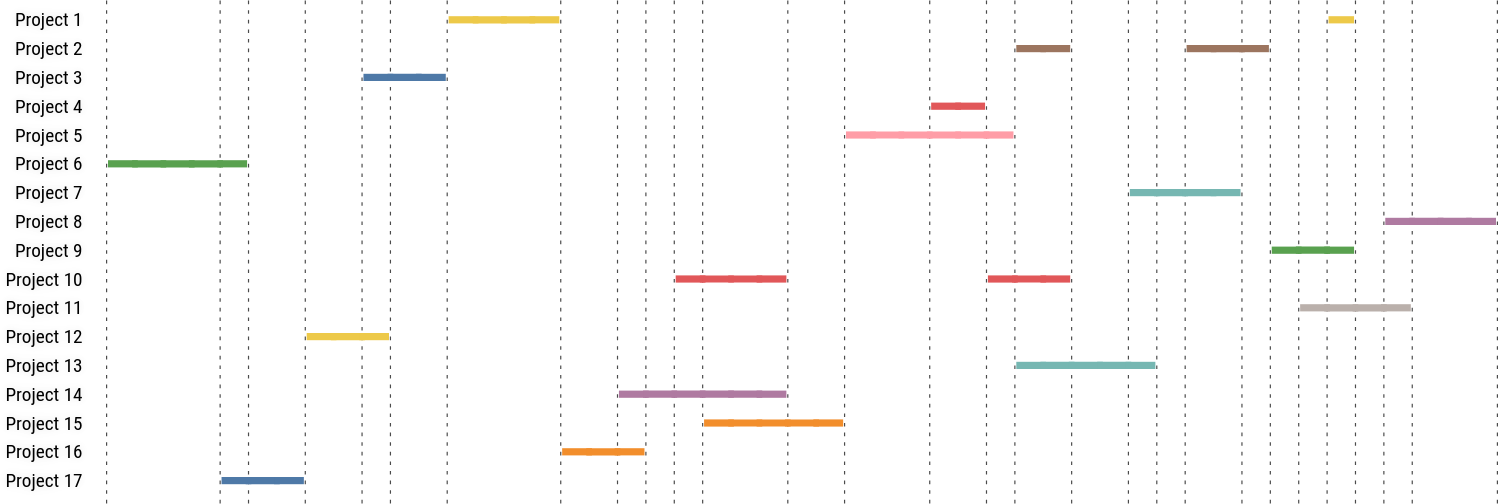}}
  \end{minipage}}
\usebox{\measurebox}\qquad
\hfill
\begin{minipage}[b][\ht\measurebox][s]{.46\textwidth}
\centering
\subfloat
  [LinSets.zip with block links]
  {\label{fig:figB}\includegraphics[width=\textwidth]{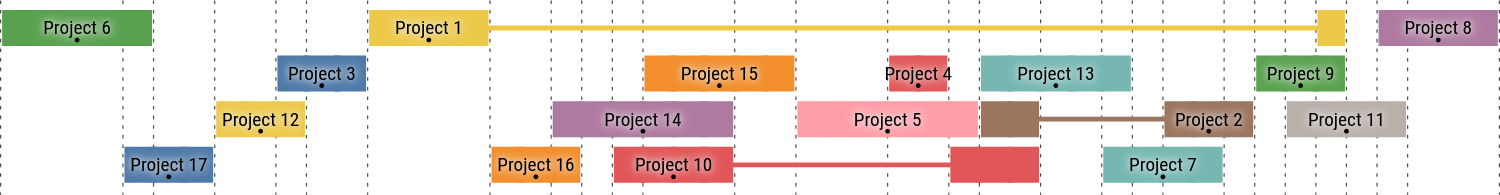}}

\vfill

\subfloat
  [LinSets.zip with block links]
  {\label{fig:figC}\includegraphics[width=\textwidth]{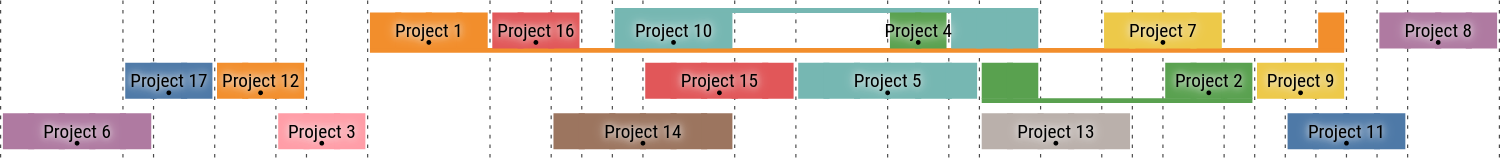}}
\end{minipage}
  \caption{A linear diagram (a) of a project management dataset. While the linear diagram uses vertical space very liberally, both variants (b) and (c) of the LinSet.zip approach present a more compact representation.}
  \label{fig:teaser}
\end{figure*}

\looseness=-1
One observation about linear diagrams is that they liberally use vertical, and to a lesser degree, horizontal space as each set occupies its own individual row in the matrix.  In the case of vertical space, set-to-set relationship tasks become harder for sets that are positioned in rows that are far apart. Similarly, as labels are usually positioned above the diagram, element-to-set relationships become harder with increasing distance between sets and element labels.
In case of horizontal space, sets containing only a small number of elements consequently produce lots of horizontal white space.
Therefore, in cases where the available screen size is restricted linear diagrams might be a suboptimal choice as either the diagram must be scaled down or the dataset can not be shown at the same time.

Our alternative approach LinSets.zip presents space-efficient linear set diagrams while still preserving as many design principles of linear diagrams; see \autoref{fig:teaser}. As it is considered best practice in linear diagrams, our approach also optimizes the number of line segments necessary to represent each set. These line segments resemble \emph{blocks} in our visualization due to the necessary label placement inside these blocks. Hence, we only use the term blocks throughout the paper even when specifically talking about linear diagrams.
In LinSets.zip vertical space is used more efficiently by allowing multiple compatible sets (i.e., not sharing any elements) to occupy the same row, using different colors to distinguish them. 
\revision{This works especially well when a set system has many non-overlapping sets.}
We formulate this problem as a graph coloring problem that gives rise to multiple, more restrictive, formulations. In the base variant the goal is to maximally reduce the number of rows without violating the compatibility. Other variations restrict the possibility of blocks of different sets alternating or restrict the number of alternating sets to two. All mentioned variations can also be bounded on the total number of sets that can be placed in each row.

We present several visualization styles that utilize the different variants while still being similar to linear diagrams. We introduce the concept of \emph{block links}, thin lines that connect different blocks of the same set, which aid the viewer in distinguishing between different sets in the same row.

\looseness=-1
As the underlying column ordering and graph coloring problems are computationally hard, we implemented LinSets.zip using both heuristics and exact algorithms that yield optimal solutions. \revision{The source code of the implementation is available on OSF.\footnote{\href{https://osf.io/2zwec/}{https://osf.io/2zwec} }} Based on this, we performed a quantitative analysis on runtime and quality criteria to give an indication of what problem sizes can be efficiently solved optimally and the trade-off of using fast heuristics compared to optimal algorithms.

Furthermore, we also conducted a small-scale user study to give indication if placing several sets in the same row affects a user in performing typical set visualization tasks. Here, we asked the participants to perform five standard set- and element-based tasks~\cite{AlsallakhMAHMR16} on static images generated with four variants of our framework. We measured task completion time and accuracy.

\section{Related Work}

\revision{Set visualization is a subfield of information visualization that has been extensively investigated.}
Often, set visualization is integrated in visualization systems to provide a supplementary view on the presented data. We focus mainly on visualizations that consider only abstract set data, where no additional attributes in the data are necessary. 

\looseness=-1
Alsallakh et al.~\cite{AlsallakhMAHMR16} published a state-of-the-art report on set visualization in 2016. They provide an overview of existing set visualization techniques and approaches and classify them by visual representation, scalability, and capacity of solving set-theoretic tasks according to a taxonomy. 
The taxonomy itself gives an overview over set-theoretic tasks that commonly arise in set visualization systems. The tasks are classified in three categories: \textit{element-based tasks} are tasks concerned with relationships between elements and sets; \textit{set-based tasks} are tasks concerned with relationships between two or more sets; \textit{attribute-based tasks} are concerned with the relationship between element attributes and their appearance and distribution in sets. Overall, they define a total of 26 tasks and six categories of set visualizations. In the related work we will focus on the three most relevant categories. 

Generally, Euler and Venn type diagrams are the most intuitive. In both variants sets are depicted as closed curves where overlapping regions represent set intersections. In Venn diagrams all possible intersections are shown, even if they are empty. Euler diagrams  show only non-empty set intersections. 
\revision{For Euler and Venn diagrams automated approaches with regular shapes (e.g.~\cite{BaimagambetovHS18, KehlbeckGWD22}) 
as well as irregular shapes (e.g. ~\cite{RicheD10, SimonettoAA09 ,RottmannWBGNH23})
have been proposed.}
Since well-formed Euler and Venn diagrams do not always exist, the resulting visualization might not be well-matched, i.e., non-existing intersections are shown or existing intersections are shown twice. Due to this observation, Euler diagrams hardly scale for datasets beyond 4--6 sets~\cite{AlsallakhMAHMR16}.

Matrix-based techniques are another class of techniques. Here, sets and elements are depicted as rows and columns of a matrix. Set membership of elements is indicated by coloring or marking the respective cell with a glyph in the matrix. Typically, approaches in this class are designed for interactive analysis and exploration which requires interactivity, e.g. permuting rows or columns. Examples, such as RainBio~\cite{lamy2019rainbio}, or UpSet~\cite{lex2014upset}, 
are powerful and scalable visual analytics systems that incorporate multiple views on the data. However, due to their complexity they are not necessarily intuitive. LinSets.zip is related to matrix-based techniques as the underlying visual metaphor is in its essence a matrix. Similarly, rows and columns can be permuted, however, LinSets.zip focuses on minimizing blocks by reordering columns to increase readability instead of highlighting patterns in the data. Furthermore, multiple rows of the matrix are compressed to show a more compact representation. Also, similar to linear diagrams a block in LinSets.zip represents an aggregation over multiple cells in a row instead of explicitly indicating membership of individual elements.    

Aggregation-based technique, such as Radial Sets~\cite{AlsallakhAMH13}, or \revision{MetroSets~\cite{JacobsenWKN21}} 
handle increasing number of elements by not explicitly showing individual elements belonging to sets but rather aggregating multiple set elements into one visual element to indicate membership frequency. Often, they are combined with interactivity, as in UpSet~\cite{lex2014upset}, \revision{Dual Radial Sets~\cite{MatkovicGBSH20}} or RainBio~\cite{lamy2019rainbio}, to show details of specific sets or elements on demand.  
The most relevant aggregation-based technique concerning Linset.zip are linear diagrams~\cite{RodgersSC15}. The underlying visual metaphor of linear diagrams is a matrix where each set is represented as a horizontal line in its respective row and each element as a column. 
Contrary to matrix-based techniques, contiguous cells of the matrix are represented by a single block to indicate all elements in this range belong to the set.
Several user studies have been conducted~\cite{ChapmanSRMB14, LuzM19, SatoM12} where it has been shown that linear diagrams perform equally well or better than other diagram types. Similarly, different design decisions and their impact on readability and task performance have been compared~\cite{RodgersSC15, AlqadahSHC16}. Here, findings indicated that minimizing the number of blocks (called line segments there) in the linear diagram had the most effect on readability. It has been shown that reordering columns to minimize blocks is an \NP-hard problem~\cite{ChapmanSC22, DoblerN22} but heuristic~\cite{ChapmanSRMB14, chapman2021drawing, LuzM19} and exact~\cite{DoblerN22} algorithms to compute optimal solutions have been proposed. Lastly, interactivity~\cite{Chapman21} in linear diagrams has been investigated.

LinSets.zip builds on the main findings of linear diagrams. Columns are reordered to reduce blocks while the visual encoding is kept similar to what has been experimentally validated. However, space is better utilized as several sets can be compressed into the same row. Additionally, LinSets.zip gives the option to show all set elements explicitly which we think is a viable assumption for up to 50 elements.

\revision{Compressing multiple sets into a single row is not an entirely novel concept in set visualization and has been explored previously. The rainbow boxes system~\cite{LamyBCF17} allows compressing variable height rows. Moreover, timelines or Gantt charts generally allow compression; for example, as in TimeSets~\cite{NguyenXWW16}. In both examples the problem statement is different to LinSets.zip and the proposed approaches only consider greedy heuristics. We show that finding an optimal solution is possible in reasonable time.}

\section{Design Decisions}\label{sec:designdec}
Degrees of freedom in the design of a linear diagram can differentiate between two concerns. The first concern is the layout of a linear diagram as it is determined by the mapping of sets and elements to axes of a matrix and subsequently the row and column order of said matrix. While the mapping only affects the orientation of the diagram, the order of rows and columns has more drastic implications. The column order impacts the number of blocks necessary to represent a set. For example, if elements $a$ and $c$ are in the same set then the order $(a,b,c)$ will require two blocks while the order $(a,c,b)$ only requires one. Moreover, the row order determines the visual distance between sets. The second concern is how the data is encoded as graphical features. Here, thickness of blocks, color of blocks, label placement for elements and sets, guide-lines to labels, and margins all impact the appearance of the diagram. 

From a theoretical perspective on set visualization, linear diagrams scale well with an increasing number of sets as all set-to-set relationships can be visualized. However, the height of a linear diagram is proportional to the number of sets in the set system as each row represents exactly one set. We started our investigation based on this observation with the assumption in mind that vertical space is limited. For example, static diagrams in print media, multi-view visualization interfaces, or even interactive single-view visualization interfaces all have constraints on what can be shown or perceived at the same time.

Furthermore, tasks where blocks of different sets are vertically distant should be intuitively harder to solve than tasks where the blocks are vertically close. 
Even though this could be tackled with integrating interactivity to allow reordering of rows, this could over-complicate the interface and is a non-viable solution for static visualizations.

\begin{figure}
    \centering
    \subfloat[Linear Diagram\label{fig:variant0}]{\includegraphics[width=0.35\linewidth, page=1]{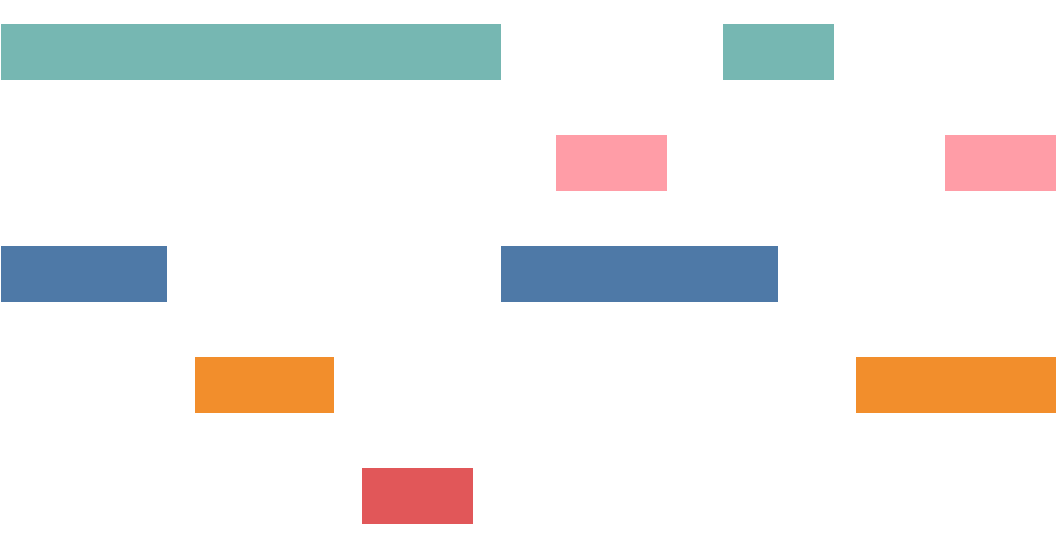}}
    \hspace{1 cm}
    \subfloat[\comp{1}\label{fig:variant1}]{\includegraphics[width=0.35\linewidth, page=2]{figures/variants.pdf}}
    \hfill
    \subfloat[\comp{2}\label{fig:variant2}]{\includegraphics[width=0.35\linewidth, page=3]{figures/variants.pdf}}
    \hspace{1 cm}
    \subfloat[\comp{3}\label{fig:variant3}]{\includegraphics[width=0.35\linewidth, page=4]{figures/variants.pdf}}
  \caption{Different visual encodings of LinSets.zip compared to linear diagrams (a). \comp{1} (c) allows non-intersecting sets in the same row. \comp{2} (b) additionally requires that blocks do not alternate. \comp{3} (d) allows for a maximum of two sets to have alternating blocks. }
  \label{fig:variants}
\end{figure}

From those observations our initial idea was to find a more space-efficient representation of linear diagrams while trying to improve or at least keep a similar level of readability. %
Our approach computes a layout that reduces the vertical space necessary to draw a linear diagram by packing multiple \emph{compatible} sets into the same row. First, we state the three definitions of compatibility that result in different visual encodings as seen in \autoref{fig:variants}. We say that two sets $A$ and $B$ \emph{alternate} if their blocks alternate under a given column order, i.e., a block of $A$ is followed by a block of $B$, which again is followed by a block of $A$ (or vice versa).

\begin{itemize}
\setlength\itemsep{0pt}
\item[$\boldsymbol{\Gamma}_{\boldsymbol{1}}$] Two sets are compatible if they do not intersect.
\item[$\boldsymbol{\Gamma}_{\boldsymbol{2}}$] Two sets are compatible if they do not intersect and their blocks do not alternate within a given column order.
\item[$\boldsymbol{\Gamma}_{\boldsymbol{3}}$] Two sets are compatible if they do not intersect and their blocks alternate only with each other within a given column order.
\end{itemize}

Compatibility definition \comp{1} is the minimum requirement, which in turn allows to maximally compress the linear diagram. However, in this case we can only distinguish between blocks of different sets in the same row by color. This assumption is problematic as humans cannot reliably distinguish between multiple colors and it is not inclusive for people with color vision deficiencies. 

To alleviate this problem we propose the refined compatibility definitions \comp{2} and \comp{3}. Here, we further restrict sets to be put in the same row, based on the occurrence of alternating blocks. The reason here is that we can use additional visual elements to indicate blocks belonging to the same set and therefore redundantly encode blocks belonging to the same set. We call this concept \emph{block links}, which are essentially thin straight lines connecting all blocks of a set as seen in \autoref{fig:variant2}. For \comp{2} we can draw the block links in the vertical center to indicate blocks belonging to the same set as blocks of different sets never alternate. In the case of \comp{3} we can draw the block links at either the top or bottom of the row and therefore we can only allow blocks of two sets to alternate at the same time. Furthermore, maximally compressing the rows can have the effect of creating diagrams that are visually too dense. We propose the option to limit the number of sets that can be placed in the same row by a positive integer $B$. 

While we have introduced the intention of compressing several sets into the same row to create more compact layouts, we have not talked much about what is considered best-practice for linear diagrams and how this ties into our approach.
We looked at existing design principles of \revision{linear diagrams} which have been empirically evaluated. Three design principles had statistically significant positive impact on set-theoretic tasks~\cite{RodgersSC15}.

\begin{itemize}
\setlength\itemsep{0pt}
 \item \textbf{Design Principle 1:} draw linear diagrams with a minimal number of blocks.
 \item \textbf{Design Principle 2:} draw linear diagrams with guide-lines at the beginning and end
 \item \textbf{Design Principle 3:} draw linear diagrams with thin blocks.
\end{itemize}

The first design principle still applies to our approach. Therefore, we try to minimize the number of blocks for all sets by reordering the columns. In the case of \comp{1} compressing sets is independent of column order. For \comp{2} and \comp{3} this is not the case as the column order is directly connected to whether blocks alternate or not. Here, we propose to first order the columns to minimize the number of blocks before compressing the linear diagram as this should decrease the visual complexity. The second design principle states that using guide-lines, thin unobtrusive vertical lines, as a visual aid to indicate the beginning and end of blocks. This can be applied to our design as it does not interfere with compression. Adhering to the third design principle is more difficult, if not impossible. As our approach packs multiple sets into the same row, it is impossible to place a single set label at the beginning of each row. Therefore,  we have to place labels near one of the blocks of each set. Here, we think that increasing the line height for representing blocks in order to embed the labels within blocks helps in identifying which label belongs to which set. Therefore, we cannot use thin lines to represent sets.

After implementing a first prototype we conducted an expert interview with a graphic designer. The designer provided valuable feedback and recommended to use margins between blocks in the same row and between rows, adding white background color to text labels to make them stand out and to limit the color palette. 

\section{Overview of the LinSets.zip Approach}

\begin{figure*}[th]
    \centering
    \subfloat[(I)]{\label{fig:pipeline1}\includegraphics[width=0.21\linewidth, page=1]{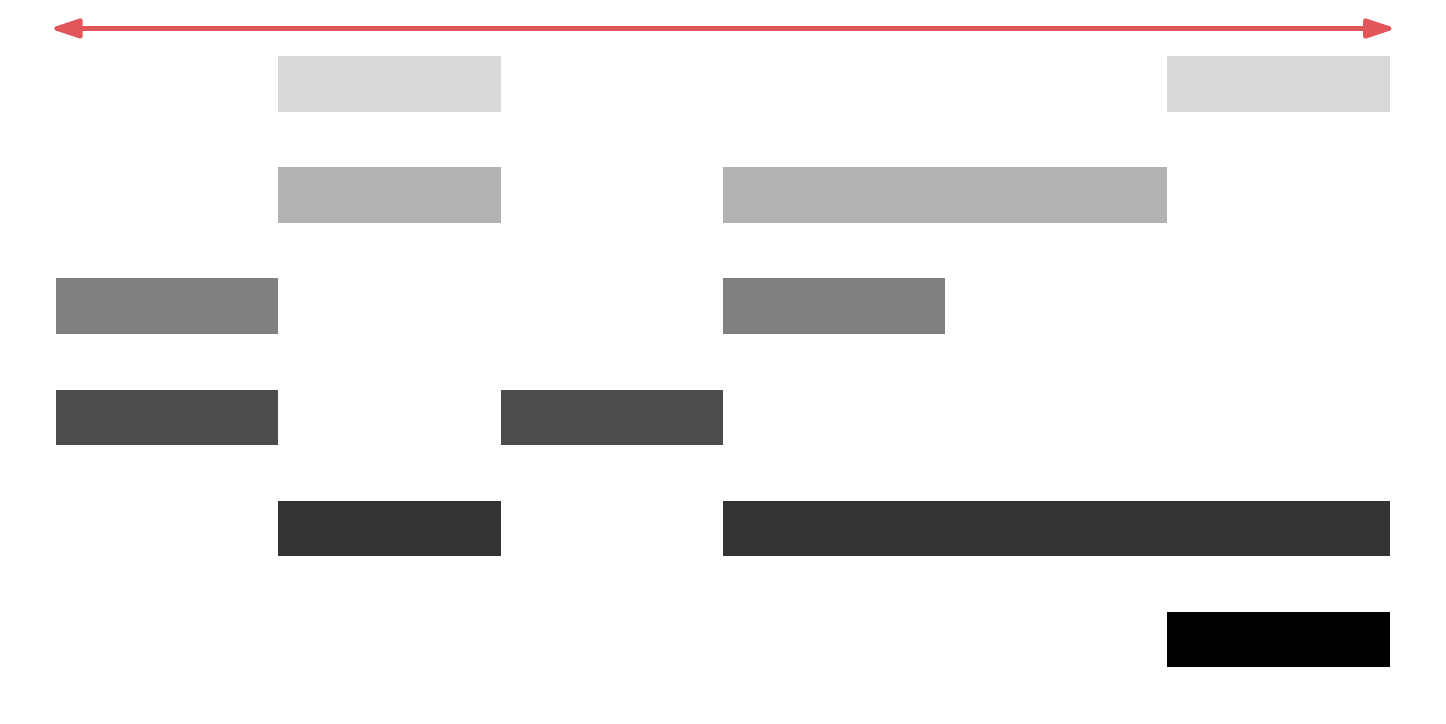}}
    \hfill
    \subfloat[(II)]{\label{fig:pipeline2}\includegraphics[width=0.21\linewidth, page=2]{figures/pipeline.pdf}}
    \hfill
    \subfloat[(III)]{\label{fig:pipeline3}\includegraphics[width=0.21\linewidth, page=3]{figures/pipeline.pdf}}
    \hfill
    \subfloat[(IV)]{\label{fig:pipeline4}\includegraphics[width=0.21\linewidth, page=4]{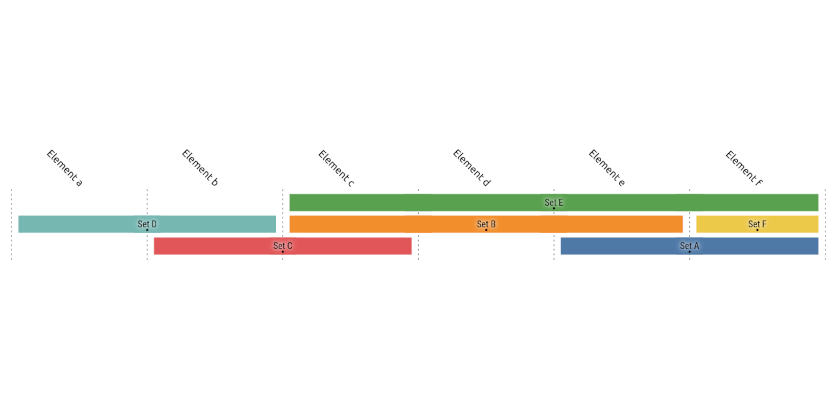}}
  \caption{First, the LinSets.zip approach reorders elements (a) of the input before a compression variant is applied (b). Next, rows are reordered and colors are assigned to sets (c). Finally, the diagram is rendered in the LinSets.zip style (d).}
  \label{fig:pipeline}
\end{figure*}

In this section we give a high level overview of the LinSets.zip approach, which is modelled as a four-stage (I--IV) pipeline depicted in \autoref{fig:pipeline}. We provide different modules for each pipeline stage.

The input is an abstract set system modelled as a hypergraph $\HG=(\V,\Sets)$. The vertex set $\V$ represents all elements in the set system and $\Sets \subseteq 2^{\V}$ the sets themselves. A detailed description of the algorithms and techniques used in stages (I) and (II) can be found in \autoref{sec:alg}.

\textbf{Column Ordering (I).} In the first stage we reorder columns in LinSets.zip such that the total number of blocks is minimized. This is equivalent with reordering columns in linear diagrams and the same techniques can be applied.
Unfortunately, this problem is \NP-hard but we can reformulate the problem of finding an optimal column order as a traveling salesperson problem. We provide modules that solve this problem optimally using an efficient travelling salesperson solver or use heuristics. The output of this stage is an ordering $\pi^c$ of the columns.

\textbf{Row Compression (II).} Next, we are concerned with minimizing the total height of the LinSets.zip diagram by compressing multiple compatible sets into the same row. To reiterate from \autoref{sec:designdec}, we present three definitions of set compatibility that each have implications for the rendering later. The first variant \comp{1} can assign sets to the same row if they do not pairwise intersect. This can be modelled as a conflict graph where each vertex represents a set and edges represent conflicts, namely  intersections between sets. By solving a graph coloring problem on the conflict graph a mapping of sets to rows can be extracted as each color represents a row in the matrix. In the second variant \comp{2} we additionally restrict that blocks of sets in the same row cannot alternate. Similarly, the third variant \comp{3} restricts that blocks of a maximum of two sets in a row can alternate at the same time. Both variants add constraints to the coloring problem. While \comp{1} works independent of a given column order, it is necessary to have a given column order for \comp{2} and \comp{3} as the definition of alternating is entirely dependent on the order of columns. 

\looseness=-1
Moreover, in some cases it might be beneficial to restrict the maximum number of sets per row, e.g., to leave space for labels or guarantee unique colors from a given palette. This can be modelled as a bounded graph coloring problem. Again, graph coloring problems are considered \revision{\NP-hard~\cite{GareyJ79}} but for both, unbounded and bounded coloring problem, we provide modules that compute an optimal solution with a minimal number of rows necessary and a heuristic solution. The output of this stage is a mapping of sets to rows.

\textbf{Row Order and Color Assignment (III).} After computing a column order and row mapping we have to consider two more aspects that impact the LinSets.zip diagram. Firstly, the row order is not fixed. Potentially, we can apply the same procedure as for column ordering to reduce vertical blocks --- consecutive rows of a column that all contain a block. However, compared to horizontal blocks it is unclear what implications this additional step has on readability. 
\revision{As row order has no impact on compression and the instances in the user experiment were small, we opted to use a random order.}

Secondly, the assignment of colors is non-trivial as the total number of sets can exceed the number of colors available in our palette. No unique color should be assigned to two sets in the same row. Furthermore, the perceptual distance between colors assigned to a row should be maximal. However, this problem is known as the maximum differential coloring problem~\cite{HuKV10} and is known to be \NP-hard. Runtime experiments on a prototype showed, unlike (I) and (II), that solving the problem optimally would bottleneck the overall runtime of the pipeline. A heuristic solution did run faster but did only marginally improve the result from a more simplistic approach. Therefore, we implemented a circular color assignment that assumes that the row order is fixed. 
Both problems open interesting questions on their own, however, due to space limitations we will not give a more comprehensive description.

\textbf{Rendering (IV).} Finally, we need to render the output by using the results computed in the previous stages (I-III). In total, we provide four rendering styles that capture linear diagrams and the three compatibility definitions of LinSets.zip. All styles are built on the visual metaphor of a matrix. Rows are used to represent (multiple) sets and columns represent the individual elements. Blocks are drawn as horizontal bars covering the respective set elements. The blocks are colored with one of the available colors in the Tableau10 color palette. For all styles we place labels for elements or the intersection cardinality above the matrix. Vertical guide-lines indicate the beginning and end of intersections. For linear diagrams we place set labels to the left of the matrix while for the LinSets.zip styles labels are placed in the largest block of their respective sets. For \comp{1} only the color is used to distinguish blocks of different sets in the same row. For \comp{2} block links are drawn in the center from the first to the last block of a given column order, or, at the top or bottom of blocks for \comp{3} when blocks of two sets are allowed to alternate. We use whitespace as margin between blocks in the same row and between rows. 

\section{Algorithms}\label{sec:alg}
In this section we go into detail about the different \textit{\textcolor{blue}{exact}} and \textit{\textcolor{purple}{heuristic}} algorithms used in the modules for stage I-II of the pipeline. 

\subsection{Column Ordering (Step I)}\label{sec:algcolorder}
\looseness=-1
We minimize the number of drawn blocks by computing a column ordering $\pi^c$.
This is done by formulating the problem as a problem on binary matrices and using a known travelling salesperson (TSP) formulation of this problem.
For the input hypergraph $\HG$ we compute the binary matrix $A$ of size $|\Sets|\times |\V|$ such that $A_{i,j}=1$ if vertex $v_j$ is in the hyperedge $E_j$. We find a column ordering of $A$ that minimizes the number of so-called \emph{blocks of consecutive ones}, which are maximal consecutive entries of ones in a row of a matrix. There is a one-to-one correspondence between these blocks of ones and the blocks in the linear diagram. 
The corresponding minimization problem is known as \emph{Consecutive Block Minimization} and is \NP-hard \cite{kouPolynomialCompleteConsecutive1977}. However, the problem can be formulated as a TSP problem \cite{haddadiConsecutiveBlockMinimization2008}:
First we add an auxiliary column of ones to the matrix $A$. Then, we construct from $A$ a graph $G$ such that vertices of $G$ correspond to columns of $A$. The distance $D_{i,j}$ between two vertices $v_i$ and $v_j$ in $G$ is the \emph{Hamming distance} $\sum_{1\le k\le |\Sets|}|A_{k,i}-A_{k,j}|$ between the corresponding columns $c_i$ and $c_j$. The auxiliary vertex $v$ corresponding to the added column of ones serves as start and end vertex of the tour, and from a \revision{minimum-length tour} in $G$ 
we obtain a permutation of the columns of $A$ with the minimum number of blocks of consecutive ones (blocks in the linear diagram), where the auxiliary column is the last one in the permutation. By removing the auxiliary column from this permutation, we obtain a permutation of the original matrix, that serves as permutation of the set $\V$. 

\textit{\textcolor{blue}{Exact.}} To find a tour of minimal distance in $G$ in the exact pipeline we use the Concorde TSP solver\footnote{\url{https://www.math.uwaterloo.ca/tsp/concorde.html}}.

\textit{\textcolor{purple}{Heuristic.}} The heuristic version applies the simulated annealing algorithm of NetworkX to find a short TSP tour, starting with an approximation based on the Christofides algorithm \cite{christofidesWorstCaseAnalysisNew2022}.

\subsection{Compression (Step II)}\label{sec:algcompression}
\looseness=-1
Once we have an ordering $\pi^c$ of the columns (the vertices \V\ of \HG), we perform the actual compression of the linear diagram.
Our aim is to use as few rows as possible.
For now, let us assume that we do not want to connect blocks of a set by block links. 
The main observation is that we can place two sets $S_1,S_2\in \Sets$ of \HG\ into the same row if and only if $S_1\cap S_2=\emptyset$.
This is independent of the column ordering $\pi^c$.
Hence, this can be modelled as a graph coloring problem: Let $G$ be a \emph{conflict graph} with $V(G)=\Sets$, $E(G)=\{\{S_1,S_2\}\mid S_1,S_2\in\Sets, S_1\cap S_2\ne\emptyset\}$, and $C$ be a set of colors.
A valid coloring $\col:V(G)\to C$ of $G$, that is $\col(u)\ne \col(v)$ for $\{u,v\}\in E(G)$, immediately gives us a compression of the linear diagram into $|C|$ rows. 
Each color $c\in C$ corresponds to a row in the linear diagram, and a set $S\in \Sets$ is in row $c$ if $\col(S)=c$.
We also give the option to bound the number of sets per row by a positive integer $B$. 
This translates to finding a coloring $\col$ such that for all $c\in C$, $|\col^{-1}(c)|\le B$.
The problem of finding such a coloring is know as \emph{Bounded Vertex Coloring} (BVC) and has been studied in the literature (\revision{refer to \cite{malagutiSurveyVertexColoring2010} for a survey}).

For both the unbounded case and the bounded case we apply a heuristic and an exact algorithm. Prior, we compute a large clique $K\subseteq V(G)$ using the NetworkX\footnote{\label{NetworkX}\url{https://networkx.org/}} implementation of an approximation by Boppana and Halldórsson \cite{boppanaApproximatingMaximumIndependent1992}.
Every valid coloring has more colors than $|K|$, and we will pre-specify a different color for each vertex $v\in K$.
We now present our algorithms for the two cases where the number of sets per row is unbounded or bounded.

\textbf{Unbounded coloring.} The \textit{\textcolor{purple}{heuristic}} algorithm to obtain a coloring with few colors is the NetworkX implementation of the greedy DSATUR algorithm \cite{brelazNewMethodsColor1979}: The vertices are colored one by one in decreasing order of their \emph{saturation}. The saturation of a vertex $v$ is the number of different colors the neighbors of $v$ are assigned to. Ties are broken by decreasing degree. When a vertex is colored, it is assigned to the first free color that none of its neighbors has.

\textit{\textcolor{blue}{Exact.}} For the exact unbounded coloring we use a standard ILP formulation of the coloring problem. 
Let $C=\{c_1,\dots, C_U\}$ be a set of colors, where $U$ is the number of colors required by the heuristic.
The ILP has binary variables $x_{v,c}$ for $v\in V(G)$ and $c\in C$, and binary variables $y_c$ for $c\in C$.
Variables $x_{v,c}$ should be one if and only if vertex $v$ has color $c$, variables $y_c$ should be one if and only if at least one vertex has color $c$. We obtain the following formulation.
\begin{align}
 \text{minimize:}\qquad\displaystyle\sum_{c\in C}y_c&\label{ilpcol1}\\
 \text{subject to:}\qquad\displaystyle\sum_{c\in C}x_{v,c}&=1,&v\in V\label{ilpcol2}\\
 x_{u,c}+x_{v,c}&\le y_c,&\{u,v\}\in E(G), c\in C\label{ilpcol3}\\
 y_{c_{i+1}}&\le y_{c_{i}},&i=1\dots U-1\label{ilpcol7}\\
 x_{v,c_v}&=1,&v\in K\label{ilpcol4}
\end{align}
\looseness=-1
The objective \eqref{ilpcol1} minimizes the number of required colors. 
\eqref{ilpcol2} ensures that each vertex has exactly one color. 
\eqref{ilpcol3} ensures that two adjacent vertices have different colors, while also forcing $y$-variables to take the correct value. We assume that each vertex has at least one neighbor for this constraint to work, otherwise we have to add further constraints $x_{v,c}\le y_{c}$ for each vertex $v$ without neighbors and each color $c\in C$.
\eqref{ilpcol7} reduces the search space.
For each $v\in K$, $c_v$ is a color in $C$ such that $c_u\ne c_v$ for two different $u,v\in K$.
With \eqref{ilpcol4} we fix these colors for the clique $K$.
It follows that from the values $x_{v,c}$ in an optimal solution of the ILP we obtain a bounded vertex coloring of $G$ with the minimum number of required colors, which gives an assignment of sets to rows that minimizes the number of rows. Namely, we color vertex $v$ with color $c$ if and only if $x_{v,c}=1$, or, in other words, we put set $v$ into row $c$.

\textbf{Bounded coloring.}
For $B=2$ we apply Edmonds Blossom Algorithm \cite{galilEfficientAlgorithmsFinding1986} for computing maximum matchings in the complement graph $G^c$ of $G$\revision{. The graph $G^c$ consists} of the same vertices as $G$ and $\{u,v\}\in E(G^c)$ if and only $\{u,v\}\not\in E(G)$.
A matching $M\subseteq E(G^c)$ of maximal size immediately gives us a bounded vertex coloring in $G$ with the minimum number of colors.
For each $\{x,y\}\in M$, we color $x$ and $y$ with the same color. All other pairs of different vertices have different colors.
We obtain a coloring with $|V(G)|-|M|$ colors, which is the minimum possible. \revision{If there was a coloring with less colors, then there would be a larger matching in $G^c$. As the algorithm runs in polynomial time, we do not need a heuristic for this case.}

\looseness=-1
Let us consider the case $B>2$.
Our \textit{\textcolor{purple}{heuristic}} is a slightly adapted version of the DSATUR vertex coloring heuristic from before. 
\revision{When coloring a vertex $v$, we must ensure that at most $B$ vertices use a given color in addition to the color being different from neighbors.
The new saturation of a vertex $v$ is the number of distinct colors of the union $C_1\cup C_2$, where $C_1$ are the colors of $v$'s neighbors and $C_2$ are the colors $c\in C$ such that $|\col^{-1}(c)|=B$.
Vertices are again colored in decreasing order of their saturation. The new augmented heuristic computes a bounded vertex coloring.}

\textit{\textcolor{blue}{Exact.}} The exact algorithm is an adaptation of the ILP above.
Again, the computed clique $K$ gives us a lower bound on the required number of colors.
Furthermore, the new bound $U$ on the required number of colors is computed by the heuristic algorithm for bounded vertex coloring explained before. \revision{The ILP consists of \eqref{ilpcol1}--\eqref{ilpcol4}, and we use the following constraint, which ensures that no more than $B$ vertices have the same color.}
\begin{align}
 \sum_{v\in V(G)}x_{v,c}&\le y_c\cdot B,&c\in C\label{ilpcol8}
\end{align}
The optimal solution of the ILP is transformed into an assignment of sets to rows in the same way as before.

\textbf{Considering block links.}
To be able to visualize block links in our two variants, we need to adapt the algorithms for compression.
Let us first specify at which column a block link of a set starts and where it ends.
A block link should have to cover all blocks of a set.
Hence, for set $S\in\Sets$ it starts at $s_S=\min\{i\mid \pi^c(i)\in S\}$ and ends at $e_S=\max\{i\mid \pi^c(i)\in S\}$.
For a set $S\in\Sets$ we define $\text{range}(S)=[s_S,e_S]$ as the \emph{active range} of the block link of a set.

\textit{\textcolor{purple}{Heuristic} and \textcolor{blue}{exact} \comp{2}.} Let us continue to adapt our algorithms for compatibility \comp{2}. If for two sets $S,S^\prime\in\Sets$ its block link ranges $\text{range}(S)$ and $\text{range}(S^\prime)$ overlap, then they cannot be placed in the same row. In fact, this is the only requirement, and we can model this by adding further edges to the conflict graph $G$ defined above: For each pair of different sets $S,S^\prime\in\Sets$ such that $\text{range}(S)\cap \text{range}(S^\prime)\ne \emptyset$ we add the edge $\{S,S^\prime\}$ to $G$ if it does not already exist. The algorithms are the same as described already, but instead they work with the slightly adapted conflict graph $G$.

\looseness=-1
If at most two block links can be drawn at once in a column of a row (model \comp{3}), then adapting our algorithms is not as straight-forward: Instead of considering pairs of sets, we have to consider triples of sets. Namely, let us call a triple $(S_1,S_2,S_3)\in \Sets^3$ of different sets a \emph{conflicting triple} if $\text{range}(S_1)\cap \text{range}(S_2)\cap \text{range}(S_3)\ne \emptyset$, and let $\mathcal{T}$ be the set of conflicting triples. We assume that a triple of conflicting sets only appears once in the set $\mathcal{T}$. We have to adapt our algorithms for the ILP-models, and heuristics such that a conflicting triple is never assigned the same color (same row).

\looseness=-1
\textit{\textcolor{purple}{Heuristic \comp{3}}.}
We extend the heuristic algorithms based on the DSATUR greedy algorithm, by first redefining the saturation of a vertex.
The saturation of an uncolored vertex $v$ is the cardinality of the set $C_v$, that consists of the colors of neighbors of $v$ and the colors $c$ such that there exists a triple $t\in\mathcal{T}$ with $v\in t$ and the other two vertices in $t$ are colored with color $c$. In the case of bounded vertex coloring, $C_v$ additionally contains the colors $c$ with $|\col^{-1}(c)|=B$.
The vertices are colored in decreasing order of their saturation, ties between sets $S$ (the vertices of $G$) are broken by decreasing values $\text{degree}_G(S)+|\{t\in \mathcal{T}\mid S\in t\}|$.
When a vertex $v$ is colored, it is assigned the first color $c$ that is not present in $C_v$.

\looseness=-1
\textit{\textcolor{blue}{Exact \comp{3}}.}
For the ILP it would be easy to model the constraints on triples. But we can reduce the number of constraints by realizing that we can encapsulate constraints imposed by multiple triples into a single constraint. For each $S\in \Sets$ let $T_S=\{S^\prime\in \Sets\mid s_S\in \text{range}(S^\prime)\}$. If $|T_S|\ge 3$ we add the following constraint to the ILP formulation, which does not allow more than 2 sets in $T_S$ to be part of a color $c$:
\begin{align}
    \sum_{S^\prime\in T_S}x_{S^\prime,c}&\le 2,&c\in C,S\in \Sets\label{ilpcol9}
\end{align}
\revision{This captures all conflicting triples, as every conflicting triple is a subset of some $T_S$.}
In this way we have at most $|\Sets|$ constraints instead of $|\Sets|^3$. The sets $T_S$ can be computed by iterating over the values $s_S$ and $e_S$ in increasing order and $T_S\ne T_{S^\prime}$ for $S\ne S^\prime$. Each constraint imposed by the triples $\mathcal{T}$ is modelled by at least one constraint of the form \eqref{ilpcol9}.

\section{Quantitative Evaluation}\label{sec:qualitativeEval}

We conducted a computational experiment with real-world data to be able to answer the following three questions in this section.
\begin{itemize}
\setlength\itemsep{0pt}
    \item[\textbf{A}:] What is the scalability with respect to runtime of our exact algorithms?
    \item[\textbf{B}:] How do our heuristic approaches compare to the exact approaches with respect to the quality metrics number of blocks (step~I) and compression (step~II)?
    \item[\textbf{C}:] By how much can we compress linear diagrams for real-world instances?
\end{itemize}

\subsection{Experimental Setup}
\looseness=-1
\textbf{Computational environment.}
All experiments were performed on a cluster of three nodes. Each node is equipped with two AMD EPYC 7402, 2.80GHz 24-core processors and 1 TB of RAM. All implementations were done in Python 3.7. The ILP formulations given in \autoref{sec:alg} were optimized using the Gurobi\footnote{\url{https://www.gurobi.com/}} optimizer. \revision{Multithreading was disabled, so the algorithms will perform similarily on end-user hardware whose processors have similar or higher processor frequency.}

\looseness=-1
For each instance in our dataset we performed 12 experiments for combinations of our algorithmic pipeline. We performed experiments for the heuristic-based and exact pipelines. 
Each experiment consists of three steps, corresponding to steps in the pipeline: column ordering, compression, and row ordering. The rendering step (IV) has no influence on this experiment. In our evaluation we prioritise the first two steps, as the row ordering algorithm is equivalent to the column ordering algorithm, and column ordering is evidently more important~\cite{RodgersSC15}.
In a heuristic experiment all of these steps are performed by the heuristic, hence, the output of one step is the input for the next step. In an exact experiment, all of these steps are performed by exact algorithms, unless they run into the pre-specified timeout of 300s---in that case the heuristic solution is used for the remaining pipeline steps.
For the compression we consider six options: First, we evaluate the three compatibilities $\comp{1}$, $\comp{2}$, and $\comp{3}$ between sets. Second, we consider the unbounded case ($B=\infty$), where any number of sets can be put into a single row, and the bounded case ($B=3$), \revision{where an upper bound of three sets per row is specified}. This results in $2\cdot 3=6$ \emph{compression variants}, and $12$ combinations overall.

\textbf{Instances.}
We performed all experiment on a set of real world-instances taken from DBLP\footnote{\url{https://dblp.org/}}.
These instances correspond to papers from the Graph Drawing conferences 1994--2021, from the PacificVis conferences 2001--2022, and the Symposium on Computational Geometry 1985--2022. Each paper corresponds to a set and each author corresponds to an element. We disregard papers, that do not have any overlapping author with another paper, as those do not influence the combinatorial complexity of an instance.
We generated instances by taking all papers from one conference and from year $x$ to year $y$ where $\text{first}\le x\le y\le\text{last}$, and $y-x\le 10$. The values first and last are the first and last year of a conference we are considering, e.g., 2001 and 2022 for the PacificVis conference.
Overall, we have 734 instances, with up to 627 sets, and up to 877 elements.

\textbf{Metrics.}
For question A, we record the runtime for each step of the pipeline. Further, we record the number of blocks resulting from the column order, and the \emph{compression ratio}, which is the ratio between the number of rows required by the compressed linear diagram and the number of sets in the instance.
\subsection{Results}
Let us now present the results of our experiments, answering each question A--C separately. 

\begin{figure}[t!]
    \centering
    \includegraphics{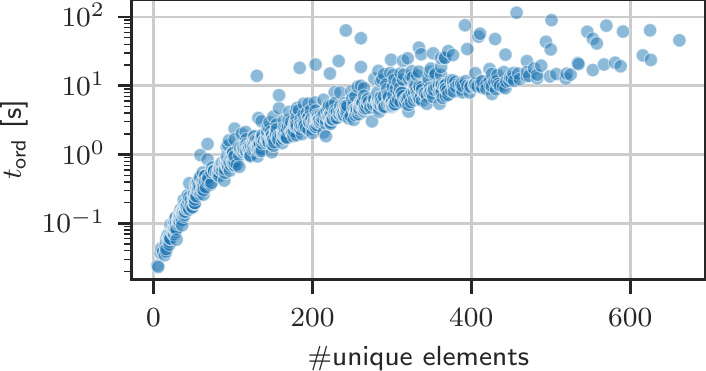}
    \caption{Time required for column ordering by number of unique elements \revision{for the exact pipeline.} \revision{The $y$-axis is scaled logarithmically.}}
    \label{fig:tcolorder}
\end{figure}

\textbf{Scalability of exact algorithms (Question A).}
We start by presenting the time required for the ordering of columns by our TSP model in the exact pipeline, denoted by $t_{\text{ord}}$. We present these values in \autoref{fig:tcolorder} by the number of unique elements---elements that are non-equivalent with regard to their set-membership, and which are directly proportional to the size of the resulting TSP-instance.
Each data point represents one instance, taking the average of five executions of our algorithm. Even though the maximum number of unique elements is 662 and the underlying computational problem is \NP-hard, only two instances exceeded the time limit of 300s. From the figure it is not immediate that the runtime increases exponentially. We suspect that in most cases the generation of the distance matrix for the TSP-solver dominates the actual time required by the solver. The outliers are instances where the TSP solver actually takes more time. Nonetheless, instances with less than 100 columns that are realistically suitable for a linear-diagram-based set visualization, all require less than two seconds.

\begin{figure}[t!]
    \centering
    \includegraphics{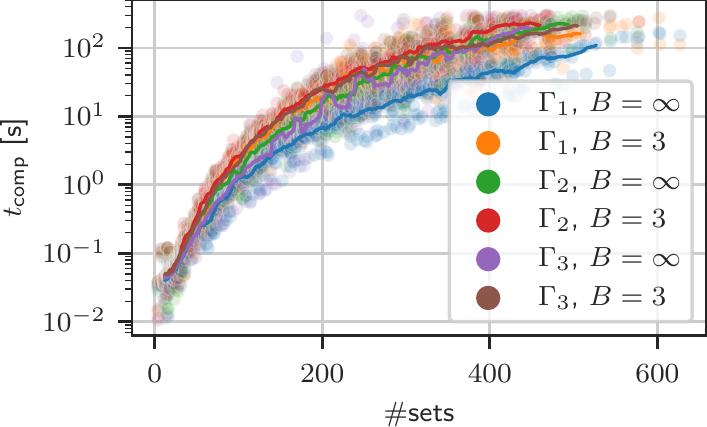}
    \caption{Time required for the compression by the number of sets in the instance \revision{for the exact pipeline.} \revision{The $y$-axis is scaled logarithmically.} The lines correspond to a running mean of 20 data points for the corresponding compression.}
    \label{fig:tcompression}
\end{figure}
\autoref{fig:tcompression} shows the runtime for the different compression variants in the exact pipeline ($t_{\text{comp}}$) by the number of sets, including a running mean.
We can clearly see that \comp{1} without bound takes the least time, while the asymptotic growth of the running time follows a similar function for all compression variants. \revision{The outliers of \comp{3} without bound result in the jagged running mean.}
Instances with less than 100 sets  all require less then 5 seconds for compression.

Let us also briefly discuss the timeouts occurring during compression, the full data can be found \revision{on  \href{https://osf.io/2zwec/?view_only=11c22b68f49a439eb3b68e372deac14f}{OSF}}. There was no timeout for \comp{1} without bound. Overall, all timeouts appeared in instances which contain over 200 sets. Each compression variants had less than 50 timeouts out of the 734 instances, the only exception being \comp{3} without bound with 70 timeouts.
To summarize this evaluation, we conclude that the exact pipeline works quite fast for all instances whose size is realistic for visualization in applications.

Detailed runtimes for the heuristics \revision{can be found on \href{https://osf.io/2zwec/?view_only=11c22b68f49a439eb3b68e372deac14f}{OSF}}. Overall, the whole heuristic pipeline took at most 125 seconds for the largest instance. The heuristics are faster than the exact algorithms, but most of the time the difference is within one order of magnitude. Thus, the exact pipeline is preferable unless dealing with hard instances that time out during execution.

\textbf{Performance of the heuristics (Question B).}
Next we want to compare the heuristic pipeline to the exact pipeline with regard to quality metrics. For this, we present for our quality metrics \emph{blocks}, and \emph{compression ratio}, the exact to heuristic ratio (EH-ratio). The EH-ratio is the value of a quality metric achieved by the exact pipeline divided by the quality metric of the heuristic pipeline on the same instance.

\revision{The EH-ratio for the number of blocks is between 1 and 1.15 for all instances but one. That means that the heuristic algorithm is no more than 15\% worse than the exact algorithm with regard to the number of blocks in all instances but one. A scatter plot of the EH-ratio for the number of blocks can be found on \href{https://osf.io/2zwec/?view_only=11c22b68f49a439eb3b68e372deac14f}{OSF}.}

\begin{figure}[t!]
    \centering
    \includegraphics{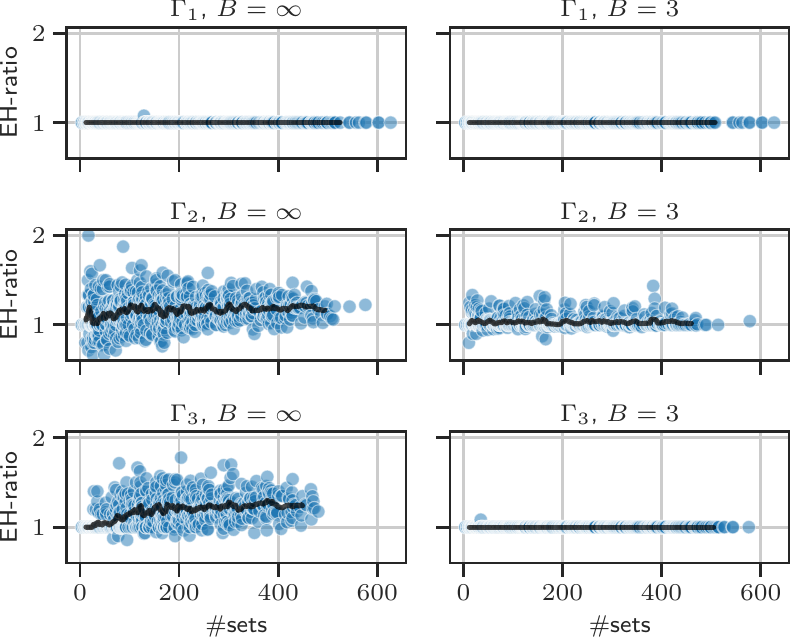}
    \caption{The EH-ratio of the compression ratio by the number of sets for the different compression variants. The lines correspond to a
running mean of 20 data points.}
    \label{fig:compcompression}
\end{figure}
\autoref{fig:compcompression} shows the EH-ratio for the compression faceted by the different compression variants. 
Values below one are possible now because models \revision{\comp{2} and \comp{3}} depend on the column order which is different for the heuristic and exact pipelines. \revision{Note that even though our pipeline uses exact modules, overall it does not necessarily compute linear diagram with maximal compression for \comp{2} and \comp{3}.}
The heuristics perform almost the same as the exact algorithms for \comp{1} for both the bounded as unbounded case. 
The heuristic for the unbounded case of \comp{3} also performs quite well, we suspect that this is because \comp{3} is not as restricting as \comp{2} and only packing a fixed amount of sets into a row makes the problem ``easier'' in less restricted coloring variants.
The largest differences between heuristics and exact pipelines are visible in the unbounded case for \revision{\comp{2} and \comp{3}}, which could have two reasons: Either different column orders have a large impact on the performance of a compression algorithm in these configurations, or our heuristics for these combinations have the potential to be further improved.
Still, the running mean line shows that in expectation the heuristic and exact pipeline perform rather similar, with the exact one being slightly better.

\textbf{Achievable compression with LinSet.zip (Question C).}
To conclude this evaluation, we  present the compression ratios that could be achieved by the exact pipeline for the different compression variants.
\begin{figure}[t!]
    \centering
    \includegraphics{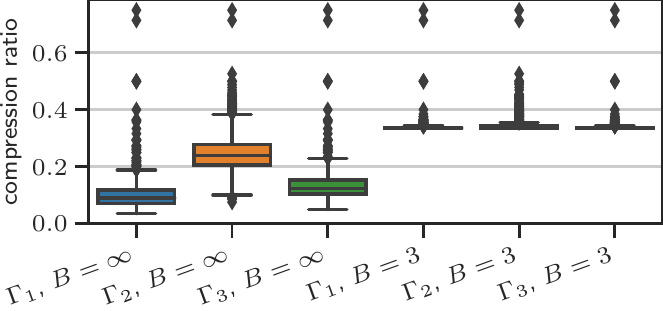}
    \caption{Boxplot of the compression ratio achieved by the exact pipeline with different compression variants.}
    \label{fig:compressratio}
\end{figure}
These are presented as boxplots with outliers in \autoref{fig:compressratio}. We see that in the bounded case the maximum achievable compression of 1/3 is achieved in almost all of the instances. 
In the unbounded case we observe the expected differences between the different models resulting from the constraints on the compression that these models impose. Namely, \comp{1} allows for the most compression, followed by \comp{3} and \comp{2}. 
Overall, LinSet.zip allows for significant compression in any variant for a large set of real-world data.

\section{User Experiment}
To better understand the impact of compressing linear diagrams on diagram readability, we conducted a small-scale user study on accuracy and task completion time over multiple tasks on static images generated with four distinct rendering styles. We compared linear diagrams (\comp{0}) to three variants of LinSets.zip that represent the different set compatibility definitions (\comp{1}, \comp{2}, and \comp{3}). Mainly, we were interested if the higher information density of LinSets.zip diagrams would lead to worse performance than linear diagrams. We carefully selected element-based and set-based tasks that spanned the spectrum of typical set visualization tasks. All datasets used were real-world instances. The raw datasets, stimuli, screenshots of the study as well as the evaluation code can be found on \href{https://osf.io/2zwec/?view_only=11c22b68f49a439eb3b68e372deac14f}{OSF}.   

\subsection{Participants and Setting}

\looseness=-1
Even though conditions in a lab setting can be better controlled, we conducted the study in an online setting. We designed the experiment as a within-group experiment where every participant was shown all five tasks on the four different styles. The target time for completion was set between 15-20 minutes. Participation required to be at least 18 years of age, no known color vision deficiency and a sufficiently large screen (at least 768px wide). In total, we were able to gather 52 complete responses. $73\%$ of the participants identified as male, $25\%$ of participants identified as female and the rest did prefer not to answer. $76\%$ of all participants were between 18-34 years old with a trivial number of participants in the remaining age brackets. Overall, the participants were well educated with all having a post-secondary degree or higher. The average self-rating of knowledge on the topic of set visualization was $2.73$ out of $5$ as most participants reported to have at least some degree of knowledge. 
We required participants to agree to not change the screen size after a screening question, which asked if they could distinguish between all visual elements and fully see the diagram, and logged their screen width. 
All participants fulfilled the requirement to have a large enough screen. 

\subsection{Datasets}
We used the same approach as in \autoref{sec:qualitativeEval} and extracted co-authorship hypergraphs of single years of the Graph Drawing conference.
From all extracted datasets we picked five that had between 36 and 56 elements and 16 to 19 sets. 
Next, we replaced set labels with arbitrary labels `Project 1' to `Project i' and assigned random, gender-balanced, names for element labels. This was necessary to ensure label readability and minimize confounding factors of labeling and familarity.    

\revision{
Each dataset was assigned to exactly one task to eliminate confounding factors of using different datasets for each style. To mitigate learning effects we randomly permuted set and element labeling to have exactly one unique task-dataset pair for each style.}

\subsection{Tasks}

We selected several tasks the participants had to complete. The selection was guided by the task taxonomy of  Alsallakh et al.~\cite{AlsallakhMAHMR16}. Below are the tasks, how the questions were stated, and the possible answers participants could give.

\begin{itemize}
\setlength\itemsep{0pt}
	\item[\textbf{T1}:] \textbf{Find sets containing a specific element:} which projects do ``Alice" and ``Bob" have in common? All projects were given as possible answers.
\item[\textbf{T2}:] \textbf{Find/Select elements that belong to a specific set:} check all people who are in ``Project i''. Six people were given as possible answers.
\item[\textbf{T3}:] \textbf{Analyze and compare set cardinalities:} how many people does ``Project i'' have in total? Ten possible values where given as possible answers
\item[\textbf{T4}:] \textbf{Analyze intersection relations:} which project(s) overlap with ``Project i''? All other projects were given as possible answers.
\item[\textbf{T5}:] \textbf{Analyze and compare intersection cardinalities:} with which project(s) does ``Project 1'' have the most overlaps in project members? All other projects were given as possible answers.
\end{itemize}

T1 and T2 are element-based tasks while T3--T5 are set-based tasks. All questions were modelled as multiple-choice questions. Whenever we asked for projects in the task we used the same order, from Project 1 -- Project X, in the answer as otherwise participants had to spend time locating their respective answer. For T2 we ordered the names in the answer as they appeared in the column order of the stimuli.  

\subsection{Stimuli}

We generated static images, or stimuli, for all pairs of datasets and styles. All images were generated with our own implementation and used the same color palette, font and font-size. We fixed the width of the generated images to 1270px but scaled down to 70\% of available screen size if a participants screen size was less. The height was solely set by the requirements of the style. To remove potential confounding factors and floor effects we paid attention to positioning visual elements representing sets at similar locations in the images. For example, we did manually move rows to be in the center instead of the bottom or top as this would have a direct impact on the visual distance between labels and the set asked in the task. Similarly, we used image processing software to place easily identifiable dots over labels in T1. The reason here is that otherwise variance in time could be explained mostly by the time a participant requires to find both elements. \revision{Adding such dots to other tasks was not required as the variance of finding sets in styles with different information density is exactly what we wanted to capture in the experiment.} In some cases for \comp{2} and \comp{3} we had overlapping labels. We ensured that no question was asked, where participants had to identify sets with overlapping labels. Lastly, we mirrored the column order for \comp{2} and \comp{3} to further mitigate potential learning effects of participants when shown the same dataset.

\subsection{Experimental Procedure} 

\looseness=-1
For each of the four conditions we asked the participants to solve two element-based tasks and three set-based tasks. In total we had $4\,\text{conditions} \, \times (2 + 3)\,\text{tasks}\,=20\, \text{trials}$. The experiment followed a five stage template of (1) consent and screening, (2) demographic questions, (3) tutorial, (4) formal study and (5) post-task questionnaire. In (1) participants were given the general study information, requirements, study procedure, data policy and consent form. 
After giving consent and agreeing to meet the requirements we showed the participants one image and asked if the image is correctly displayed on their system as well as if they are able to identify all visual elements. Also, we required participants to not change their screen size during the experiment. Afterwards in (2), we asked demographic information such as age, gender, education and prior knowledge on set visualization. Next, we showed the participants a tutorial (3) of all 5 tasks. As each task was paired with a different style the participants were familiarized with all styles and tasks in the study. We asked the participants to take their time and only proceed when they understood the task and style. Furthermore, we only allowed participants to proceed if their selected answers were correct. Next, the formal study (4) was conducted. As this was timed, we reminded the participants to answer questions correctly but as quickly as possible. The participants were given the same task on all four styles and for each participant we permuted the order of tasks and the order of styles in a specific task to minimize the learning effect. In order to reduce fatigue we showed participants a break screen between task groups which paused the timing until they proceeded. Finally, we collected qualitative information (5) about how confident they were on a 5-point Likert scale, how likely they would use the style again on a 5-point Likert scale, and a free-form text field to leave general thoughts on the study and layout styles.

\subsection{Pilot}

Before the actual experiment we invited several people to take part in a pilot study and answer a questionnaire about the study design afterwards. A total of five people with various levels of expertise in the design of user studies agreed to participate. Overall, the pilot participants did not vocalize major concerns about the general design. All participants finished the study slightly above the predicted target time. Therefore, we removed some qualitative questions at the beginning and end of the study. One participant asked us to clarify some details in the tutorial, which we implemented for the final version. We specifically asked if the participants could identify floor or ceiling effects, which all denied. Also, we asked the participants if they were aware that the underlying data and questions were the same for each style. Two participants noticed but thought that this did not help them answer the questions. Still, we adapted the final version by mirroring the element order in two of the four styles to mitigate this effect.  

\subsection{Hypotheses}

Before conducting the user experiment we formulated several alternative hypotheses. If we state that style A outperforms style B on task accuracy, this means that style A has a higher accuracy. Similarly, if we state that style A outperforms style B on task completion time then style A has a lower task completion time.

\begin{itemize}\setlength\itemsep{0pt}
    \item[\textbf{H1}:] \comp{0} will outperform all other styles on accuracy and task completion time for element-based tasks (T1, T2).
    \item[\textbf{H2}:] \comp{0} will perform worse than \comp{2} and \comp{3} on accuracy and task completion time for set-based tasks (T3-T5).
    \item[\textbf{H3}:] \comp{1} will perform worse on accuracy and task completion time than styles \comp{2} and \comp{3}.
    \item[\textbf{H4}:] \comp{2} and \comp{3} will perform similarly with no statistically significant difference.
\end{itemize}

\looseness=-1
Hypothesis H1 was guided by the intuition that element-based tasks are easier in linear diagrams as correctly identifying sets only requires scanning the vertical line below an element and finding the labels of the respective sets at its expected position. For the other styles a second visual search for the project labels is necessary. Hypothesis H2 captures the intuition that set-based tasks are easier in LinSets.zip. Here we assumed that the more compact representation makes it easier to identify the involved sets. Hypothesis H3 captures the intuition that block links make it easier to identify which blocks belong to the same set instead of solely relying on color. Lastly, the intuition in H4 is neither style with block links has an inherent advantage over the other.

\subsection{Results and Analysis}

\begin{figure}
    \centering
    \includegraphics[width=0.95\linewidth]{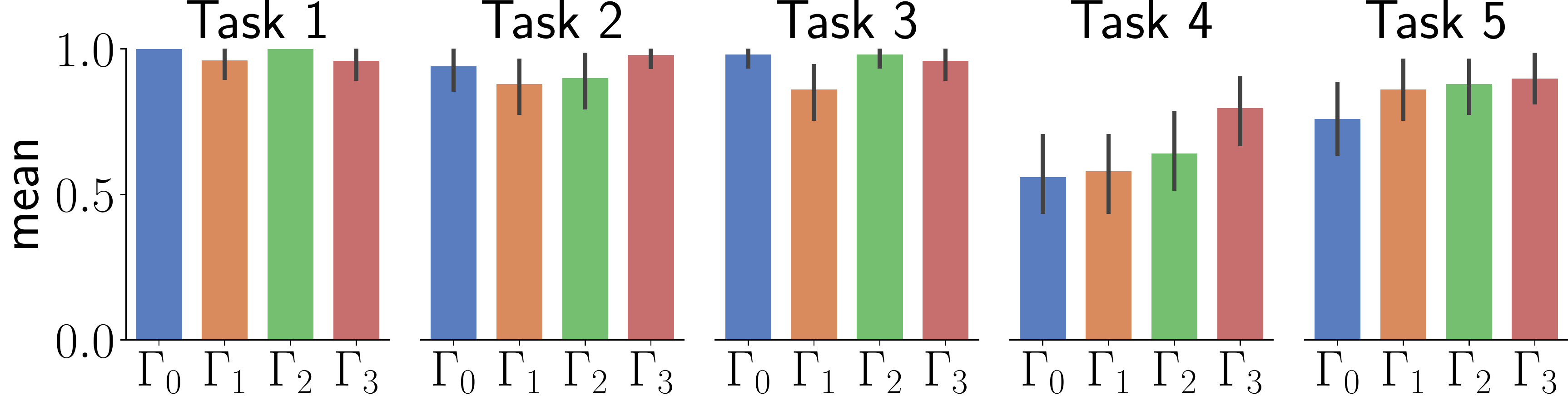}
    \caption{The participant's mean accuracy on the individual tasks.}
    \label{fig:task_perform}
\end{figure}

\looseness=-1
We describe the used statistical tests for task accuracy, task completion time and qualitative feedback below. Generally, we performed a post-hoc analysis if a statistical significance ($\alpha = 0.05$) was given. 
We evaluated each task independently and the full tables containing the statistical analysis can be found \revision{on \href{https://osf.io/2zwec/?view_only=11c22b68f49a439eb3b68e372deac14f}{OSF}}. \autoref{fig:task_perform} shows the mean task accuracy.

\textbf{Task Accuracy.} As participants' answers are binary correct/incorrect dependent variables we used Cochran's Q test to determine if there are statistical significant differences between the styles. If a significant difference was detected, we created a pair-wise contingency table and performed an asymptotic McNemar's test. 

For element-based tasks T1 and T2 and set-based task T5 we could not find statistically significant differences between any of the styles. For T3 there were statistically significant differences ($p = 0.02$). However, the pair-wise comparison only showed that \comp{0} ($p = 0.03$) and \comp{2} ($p=0.03$) outperformed \comp{1} while no significant difference was found for all other pairs. In T4 ($p=0.03$), \comp{3} outperformed \comp{0} ($p = 0.01$), \comp{1} ($p = 0.03$) and \comp{2} ($p=0.03$).

In summary, there is only marginal support for H2 and H3 and we have to reject hypotheses H1, H2 and H3 on task accuracy. H4 is supported by our findings with the exception of T4.  

\textbf{Task Completion Time.} We first tested the task completion time for normal distribution. As this was not the case we applied Friedman's test for repeated measurements with the F-Test method. Whenever we detected a statistically significant difference we performed a two-sided non-parametric pair-wise test. 

T1 showed a significant difference ($p < 0.01$) where \comp{0} is significantly outperformed by \comp{1} ($p=0.04$) and \comp{2} ($p<0.01$). \comp{2} outperforms \comp{3} ($p=0.04$). For T2 ($p=0.02$) \comp{3} outperformed \comp{0} ($p<0.01$), \comp{1} ($p=0.01$) and \comp{2} ($p=0.01$). In the case of T3 the pair-wise tests showed significant difference ($p < 0.01$) between all styles. The ranking of performance was \comp{0}, \comp{1}, \comp{2} and then \comp{3}. Significance ($p<0.01$) was also detected between the styles in T4. The pair-wise test showed that the only significant differences were between \comp{0} outperforming \comp{2} ($p<0.01$) and \comp{1} outperforming \comp{2} ($p<0.01$). For task T5 we did not detect any significant differences in completion time between the different styles.

\looseness=-1
Overall, H1 is partially supported in T1 but not in T2. H2 is unsupported while there is some partial support for H4 in T5. H3 is only partially supported in T2 and T3.  

\begin{figure}
    \centering
    \subfloat[Confidence]{\label{fig:qualitative_answers1}\includegraphics[width=0.49\linewidth]{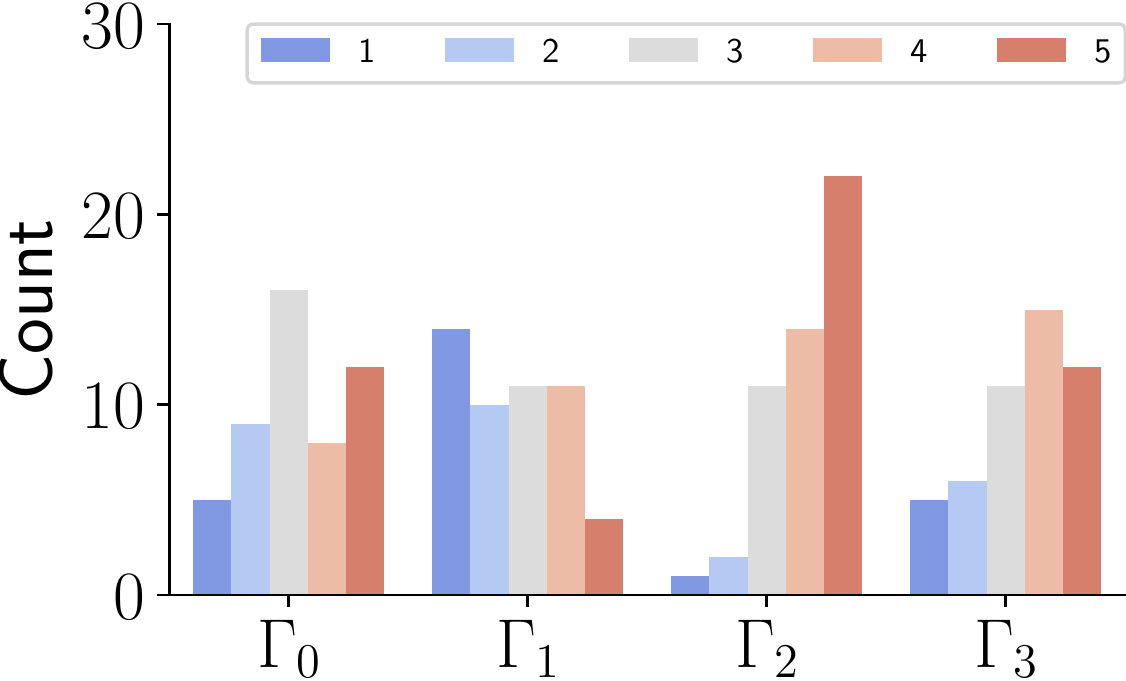}}
    \hfill
    \subfloat[Interest in using again.]{\label{fig:qualitative_answers2}\includegraphics[width=0.49\linewidth]{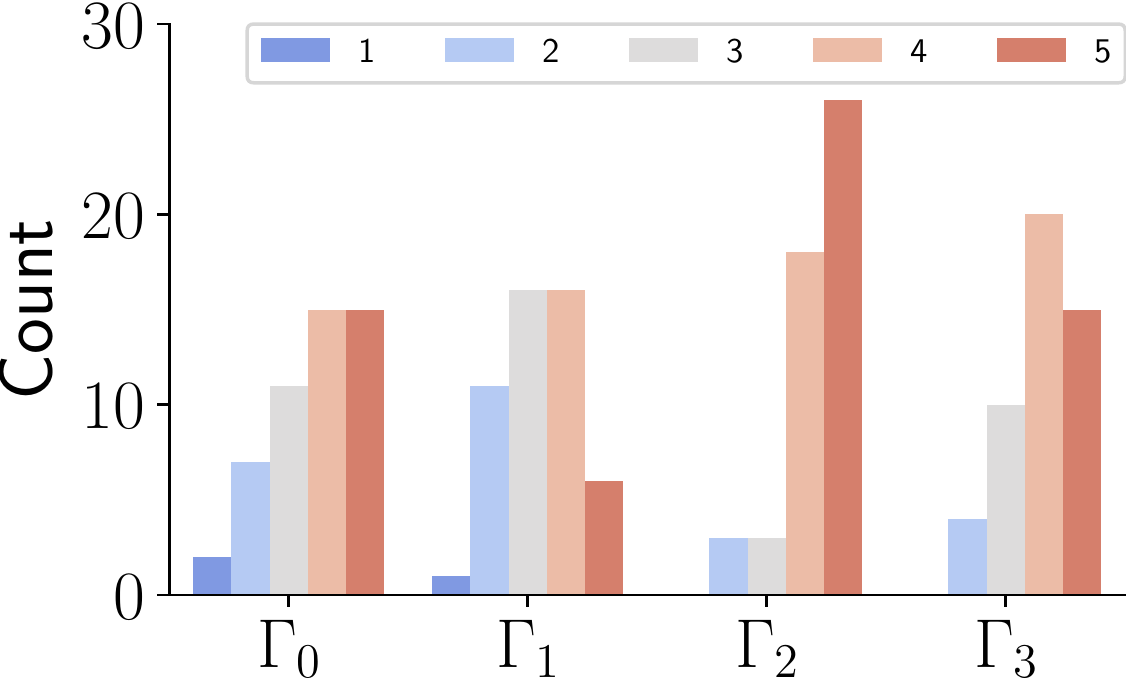}}
  \caption{5-point Likert scale (1 -- worst, 5 -- best)  of qualitative answers.}
  \label{fig:qualitative_answers}
\end{figure}

\looseness=-1
\textbf{Qualitative Feedback.} We used the Kruskal–Wallis test to identify significant differences between styles on the two qualitative answers. As both questions had significant difference, we also performed a Mann-Whitney U test with Bonferroni correction. We report the \textit{area under the curve} ($\auc$) to measure effect size.
The $\auc$ value can be interpreted as the chance that one value is larger than the other when randomly picking two samples.  Finally, we computed a Spearman correlation between performance of participants and their qualitative answers. For this we summed task accuracy and task completion time for each participant.
The distribution of participant's answers can be seen in \autoref{fig:qualitative_answers}.

There were statistically significant ($p < 0.01$) differences between the styles regarding how confident participants were about giving a correct answer. Participants felt least confident with \comp{1} which can be attributed to high information density and the fact that blocks of the same set only indicate this via color. We could not find a significant difference between \comp{0} and \comp{3}. Participants felt most confident with \comp{2} compared against \comp{0} ($\auc=0.67$), \comp{1} ($\auc=0.78$) and \comp{3} ($\auc=0.63$). We could not find statistically significant correlation between confidence and task accuracy or task completion time.    

\looseness=-1
Similarly, participant's answers if they would use a style again had significant ($p < 0.01$) differences. There is no significance between \comp{0} compared to \comp{1} or \comp{3}. Again, people felt less likely to use \comp{1} compared to \comp{2} ($\auc=0.77$) and \comp{3} ($\auc=0.68$). Participants felt most confident with \comp{2} compared against \comp{0} ($\auc=0.65$) and \comp{3} ($\auc=0.62$). Besides \comp{1} (correlation $=0.30$) on completion time, we could not find significant correlation to task accuracy or task completion time.    

We read all answers in the free-form text question and report on three general sentiments that recurred. The first recurring sentiment is that people generally agreed that \comp{1} is confusing and stated that they had difficulties either finding the correct project labels or figuring out which blocks belonged to a project. The second recurring sentiment was that participants see more benefits in linear diagrams over LinSets.zip as the visual representation is clearer and labels are always at an expected position. The third sentiment stands in contrast to the second. Here, participants stated that they had problems tracing sets to labels in linear diagrams and much more prefer the compact representation of \comp{2} and \comp{3} of LinSets.zip.

\textbf{Discussion.}
The user study we conducted to compare different variants of LinSets.zip and linear diagrams has shown that there are differences between the systems. However, we could not conclude that the more predictable and clean style of linear diagrams outperformed LinSets.zip on task accuracy and task completion time. Similarly, there is also no indication that a more compact diagram has a clear advantage over linear diagrams. On the other hand, we can draw the conclusion that LinSets.zip does perform equally well to linear diagrams and can therefore be a viable alternative when vertical space is limited. This finding is also supported by the qualitative feedback gathered in the study. Participants felt a similar level of confidence with linear diagrams and variant \comp{2} of LinSets.zip. 

\looseness=-1
Similarly, we could not find clear statistical differences between \comp{1}, \comp{2}, and \comp{3} of LinSets.zip. However, the qualitative feedback showed that participants appreciated the more intuitive block links instead of relying on color alone. Furthermore, there is a possibility that participants felt confident with \comp{2} as no two sets are alternating in a row, which leads to a less compact but therefore also less information dense diagram. Even though we implemented the possibility to restrict the maximum number of sets per row we did not test this in the user study. 

\section{Limitations and Future Work}

\looseness=-1
The participants we recruited for the user study tend to be well educated young men. Therefore, it is not clear how well our results generalize. Also, our study was conducted as an online experiment. It is not clear how attentive participants were during the study. We also only asked a small number of tasks from our participants. Overall, the user study gives some indication but is not necessarily conclusive. Another limitation is the placement of labels. Currently, we place labels in the largest block. This is problematic when the largest blocks of different sets are close which leads to overlapping labels. Furthermore, we assume that labels are short. Long labels have to be either truncated or shown on demand, as otherwise they would cover blocks and decrease readability. We also did not explore interactivity. In some cases interactivity could resolve some of the limitations of LinSets.zip. 

\textbf{Future Work.} We did not explore interactivity in this work. For \comp{1} it would be easy to interactively reorder columns to keep a single set together, as this style is independent of the column order, it is non-trivial for \comp{2} and \comp{3}. For the latter two adaptions it would be necessary to not invalidate block links. Furthermore, the user study in this work only gives an indication and a broader study would be needed to find a decisive conclusion. Lastly, a rendering style on concentric circles could be implemented. Most algorithms can easily adapt from a linear order to a circular order and the resulting diagram could have a predictable aspect ratio. 

\section{Conclusion}

\looseness=-1
In this paper we have presented LinSets.zip, a compact diagram to visualize set systems. As LinSets.zip is similar to linear diagrams we adopt concepts that have been evaluated in the context of linear diagrams. The design space we have explored focuses on creating maximally compact representations but we also present different variants that use block links as visual aids to create clearer and more readable visualizations. We show that the presented variants can be modelled as known coloring problems for which algorithms exist that work well in practice. Furthermore, we have implemented all variants and conducted computational experiments and a small-scale user study. The computational experiments show that striving for optimality is feasible in most cases and that real-world data can be significantly compressed. The findings of the user study indicate that the task accuracy and task completion time is on par with linear diagrams.

\ifCLASSOPTIONcompsoc
  \section*{Acknowledgments}
\else
  \section*{Acknowledgment}
\fi

\revision{The authors would like to thank all of the participants in our experiment, and especially the experts of our pilot study. 
This work has been funded by the Vienna Science and Technology Fund (WWTF) [10.47379/ICT19035].
}

\ifCLASSOPTIONcaptionsoff
  \newpage
\fi

\bibliographystyle{IEEEtran}
\bibliography{bibliography}

\clearpage

\end{document}